# Modeling and Simulation of COVID-19 Pandemic for Cincinnati Tri-State Area

Michael Rechtin[1], Vince Feldman[2], Sam Klare[3], Nathan Riddle[4], Rajnikant Sharma[5]

University of Cincinnati

*Abstract*—In this paper, we use SIR model to simulate the COVID-19 pandemic for Cincinnati Tri-State Area. We have built a representative population of Cincinnati that includes movements for traveling to stores, schools, workplaces, and traveling to friends houses. Using this model, we simulate the effect of quarantine, return to work, and panic buying. We show that that there will be a second wave of infections when people return to work and significant increase in number of infections when there is panic buying at stores with the announcement of the quarantine measures.

## I. Introduction

As the novel coronavirus, COVID-19, continues to sweep the world, countless models on a multitude of scales have been generated[1-6]. These models have played a crucial role in developing a response to the virus on a national scale (US [7], China[8], and Australia[9]). However, much less modeling and simulation has been explored at the community level. Motivated by a paper describing a small community model and simulation for an influenza pandemic[10], we have created an individual-based model that takes advantage of easily accessible publicly available data for the Cincinnati Tri-State region.

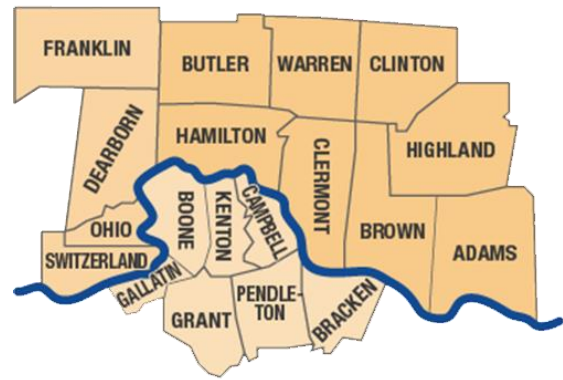

*Figure 1: Map of Counties in Cincinnati Tri-State Area*

By utilizing county specific data for households, stores, workplaces, and schools, we have been able to generate an accurate representation of the Cincinnati Tri-State Region. For our model we considered the 19 counties seen in figure 1 which constitutes a population of about 2.3 million[11]. Once these initial conditions were set, a single individual infected with COVID-19 was introduced to the population. Multiple simulations were run to assess the effectiveness of differing countermeasures. We explored implementing and releasing quarantine when varying percentages of the population were infected, as well as with varying amounts of workers being classified as essential. Besides being a community level individual-based model, our model is unique in its attention to the immediate spike in shopping when quarantine is first implemented. Since our model has been deployed relatively late, it also has more emphasis on the second wave when compared to earlier models.

This paper is an extension of a final project for our undergraduate 2020 Modeling and Simulation[12] class taught in Aerospace Engineering at the University of Cincinnati which is a part of ONR sponsored Naval workforce development program. Since our project group consists of four native Cincinnatians, we set our objective as testing the individual based modeling approach on the tri-state area. This paper describes the methodology used to create the model and the trends identified in the model's results

The rest of the paper is organized as follows. Section II presents proposed methodology for creating the population model, modeling movement, and simulation architecture. We present various simulation results in Section III and our conclusions and discussions are presented in Section IV.

[1] Michael Rechtin is an undergraduate student in Aerospace Engineering. Email: rechtima@mail.uc.edu
[2] Vince Feldman is an undergraduate student in Aerospace Engineering, Email: feldmavp@mail.uc.edu
[3] Sam Klare is an undergraduate student in Aerospace Engineering, Email: klaresh@mail.uc.edu
[4] Nathan Riddle is an undergraduate student in Aerospace Engineering, Email: riddlend@mail.uc.edu
[5] Rajnikant Sharma is an assistant professor in Aerospace Engineering, Email rajnikant.sharma@uc.edu

II. METHODOLOGY

*A. Modeling Architecture*

The model described in the paper was built using Python because to its robust support for tabular data. This capability was paramount when it came to tracking and recording the details and movements of the 23,000 individuals within the model. Using the "pandas" package, data frames were constructed to contain different layers of information about individuals and counties. The final model contained the following separate data frames: School Data, Store Data, Workplace Data, Household Data, Friends Data, and People Data. The Python script, as well as the data used in this model, can be found in our GitHub repository (https://github.com/rajnikant1010/covid19-cincinnati.git ). This model focuses on the fact that people within a community do not move randomly. People move with a purpose in the real world and in our model. Our population travels to the store, work, school, and other people's houses. We tried to capture these patterns in regards to people within the Cincinnati Tri-state area as best as we could. However, we acknowledge that there is inevitably some randomness in people's day-to-day lives. To compensate for this, some randomness was incorporated into the model through people's choices. Randomness was introduced into in our model by varying the time and day that people visit stores, work, or friends' houses.

*B. Creating Population*

To build a representative population of the Cincinnati Tri-State area, relevant data must be found. The relevance of this data is completely defined by the scope of the model, so it is critical to start with this definition. Specifically, the individual movements contained within the simulation must be defined. In the simulation described in this paper, the movements include traveling to stores, schools, workplaces, and friends' houses. This scope was defined based on the timeframe we had to create and run the model as well as available data. To begin the model, data that contained the number of schools and stores in each county was used to generate locations for schools and stores throughout each county[13-15]. The stores were split into categories of "grocery", "supercenter", and "convenience." The locations were generated using Pythons pseudorandom number generating function with a random seed. The result was a uniform spread of stores and schools throughout each county as seen in Figure 2 and Figure 3.

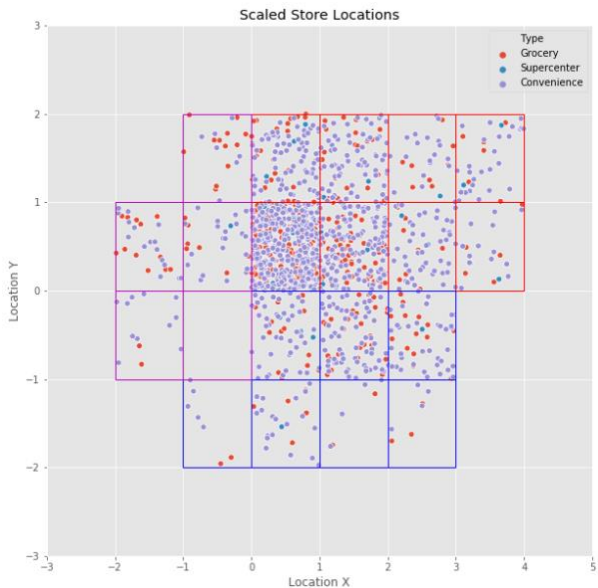
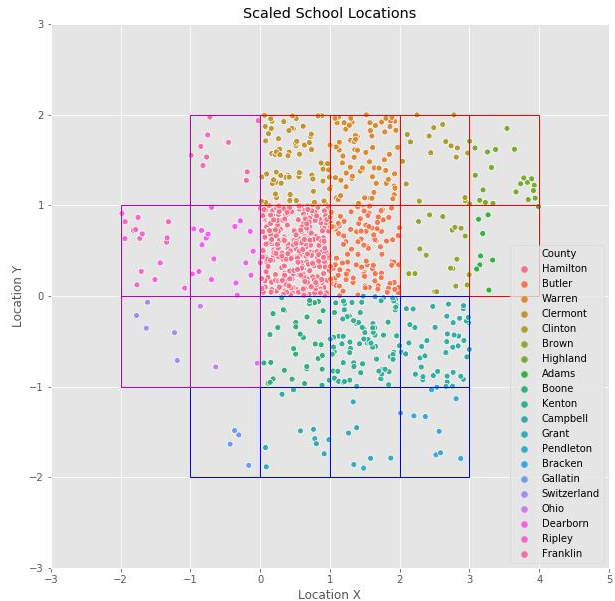

*Figure 2: Locations and types of stores in each county*

*Figure 3: Locations of simulated schools in each county*

The shape of each county was not considered for this model and they were all modeled as 1x1 squares on a plot. Rather than trying to collect, scale, and model real county specific data for workplaces, we generated an estimation based off the average commute time and size of each county. The number of workplaces in a county was meant to correctly model how quickly the virus would spread across the county due to workplace infections rather than exactly modeling the number and size of the businesses. The resulting workplaces and locations can be visualized in Figure 4. Data derived from the US Census that contains the number and average size of households in each county was used to produce a new scaled value of households in each of our simulated counties [10]. Given our simulation has a "Big O Notation" of $O(n)$, scaling the number of households by 0.01 resulted in a good balance of definition and run time. Determining the home location of each of the 9200 households was done by using the pseudorandom number generator in python to create a uniform distribution across the county. (Figure 5a) Once the households had been created and positioned, a workplace, grocery store, supercenter, and convenience store were assigned to each household based on their proximity. To account for the fact that not everyone goes to the closest store, school, or workplace to their home, some randomness was introduced into the final assignments. This introduction of randomness can be seen in the function that selects stores for each household (Figure 5b)

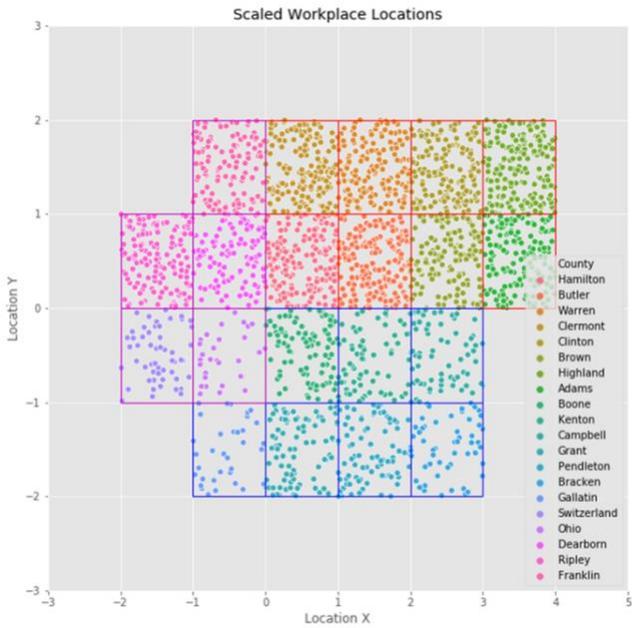

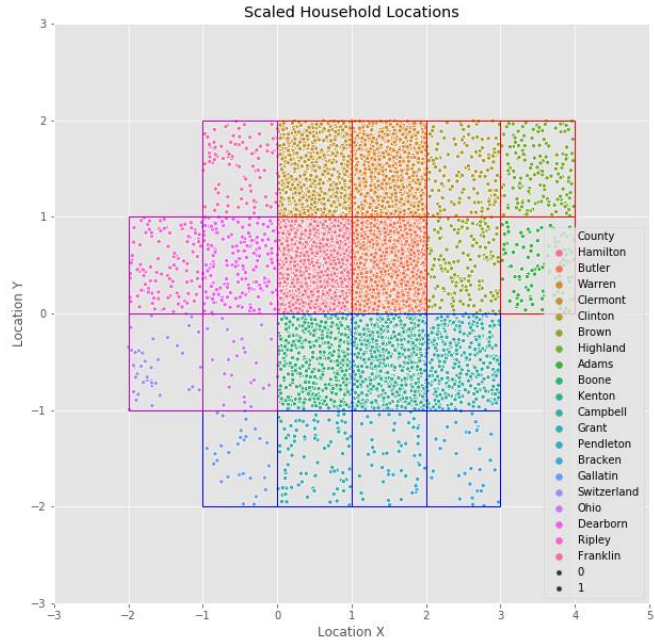

*Figure 4: Locations of simulated workplaces in each county*

*Figure 5a: Locations of simulated households in each county*

```python
def select_store(typ,xloc,yloc):
    global stores
    dist=pd.DataFrame(columns=['Store_ID','Distance'])
    stor=stores[stores['Type']==typ]
    dist['Store_ID']=stor['Store_ID']
    dist['Distance']=np.power(np.square((np.array(stor['Location X'])-xloc))
                        +np.square((np.array(stor['Location Y'])-yloc))),0.5)
    val=random.uniform(0,1)
    proximity_factor=0.9
    if val<=proximity_factor:
        store=int(dist.loc[dist['Distance']==min(dist['Distance']),'Store_ID'])
    else:
        store=random.randrange(1,stores.shape[0], 1)
    return store
```

*Figure 5b: Store selecting function chooses the closest store of a type to each household 90% of the time but a random store the remaining 10% of the time.*

An important aspect of daily life that we deemed necessary to include in our model is interactions with friends. Since the location of each household is known, a list of "friends" is built for each individual. The list includes five other households that are within a defined proximity to the original household. Throughout the simulation people of different age ranges visit their friends' houses at different times of the day. The frequency and timing these visits depend on the person age. As with the functions previously described in this paper, randomness is introduced to account for the fact that a given person's friends are not always strictly determined by proximity. Using data for the average household size in each county, the number of residents in each household was assigned accordingly [10]. Knowing the number of residents in each household allowed us to build a dataframe that contained every person from every household [10]. The resulting table is referred to as "people." The people table contains all the necessary data from the household table along with other personal information such as age and sex. Age was a key attribute of our population since it dictates daily behaviors. Eight age ranges are defined in the model, and the number of people in each age range was dictated by data collected by the US census[11]. Sex was a field within the "people" table but was ultimately not used during the simulation. The approach to building our population through a series of dataframes can be seen in Figure 6 below.

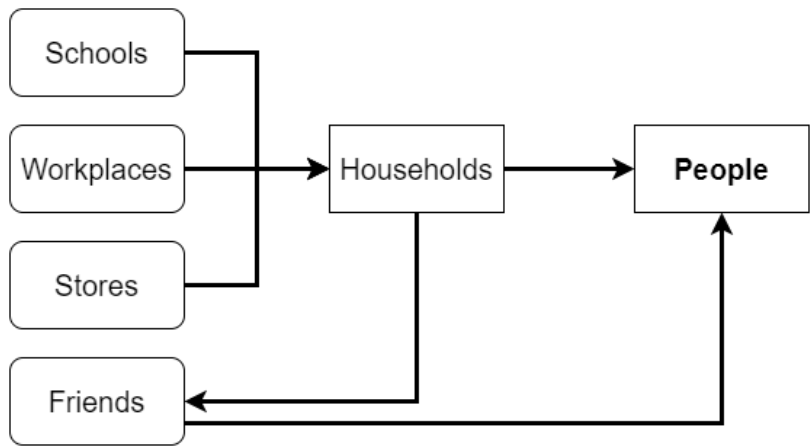

*Figure 6: Modeling approach to building a population*

*C. Modeling Movements*

    In the model described in this paper, a given day was divided into four even steps of time. Each person in the population then moves location at each of these time steps. Determining the location of each person at a given timestep was done based on age. For example, in the simulation a person below the age of 20 is located at their designated school location for the first two timesteps of the day during weekdays. In contrast, many people between 20 and 50 visit their workplace for the first two timesteps of weekdays. The full extent of the population's movements can be seen in the code posted to the GitHub repository (https://github.com/rajnikant1010/covid19-cincinnati.git ). In our model, it is assumed that everyone is home for the last timestep of each day. This simulates typical interaction with family members that can result in the transmission of the virus. People's trips to stores are random throughout the week while the frequency between their store visits is based off age. Much of the decisions for how people move throughout the duration of the simulation is based off the modelers' intuition and experience. Although a data-based approach would be preferred for obtaining the most accurate results, the approach taken was the practical choice for our particular scope. Since COVID-19 is not known to be airborne, and since travel by car is the most prevalent mode of transportation in the Cincinnati tri-state area, our model does not consider virus transmission during travel between locations.

    When quarantine is instituted in our model it is assumed that school's transition to online classes and a large percentage of people are working from home or have been temporarily laid off. The percentage of people working is varied throughout our test to show the effects of quarantine.

*D. Running Simulation*

    The starting point of the virus was a randomly selected person in the population. For all the runs displayed in this paper, the same population and initial infection point was used. The spread of the disease in our model is based off proximity. If a given person is at the same location as an infected person then their probability of contracting the virus is a function of the total number of people at their location. This equation is shown below.

$$Infection\ Probability = \frac{\lambda}{n} \quad \textit{Equation 1}$$

In this equation "n" is defined as the number of people at a given location, and "λ" is defined as factor that determines the virus's contagiousness. Analyzing our results showed that a λ of 0.25 resulted in a $R_o$ value of 1.75. Four stages of illness were defined: "Healthy", "No Symptoms", "Showing Symptoms", and "Recovered." The duration of each of these stages was based off preliminary CDC reports about the COVID-19 virus[6]. It is important to note that once a person is considered "Recovered" there is no probability of them contracting the disease again. A person's infection status influences their daily behavior. Persons showing symptoms in our model are 90% likely to stay home throughout the day.

III. RESULTS

The following results are divided into four different sets of simulations. The first simulation was run to obtain a baseline for how the virus spreads with no countermeasures. The second set of simulations measures the effectiveness of traditional countermeasure techniques such as online schooling and working from home. The third set of simulations compares a sudden return to work with a gradual return to work. The fourth and final set of simulations measures the effect of a panic buying period when quarantine measures are first introduced to the community.

*A. Covid-19 spread with no countermeasures*

For all simulations, a scaled population of 23,000 people was used. An initial simulation was conducted with no countermeasures to obtain baseline results for how the virus spread. This run, seen in Figure 7 below, resulted in a peak number of infected of approximately 14,000, or 60% of the population. Over the duration of the virus, nearly 90% of the population was infected and then "Recovered". The "Recovered" condition in our model replaces the traditional state of "Removed" used in most SIR models. This state accounts for both people who contracted the virus and have healthy recoveries as well as people who contract the virus and die. Once an individual is "Recovered," they are no longer able to transmit the virus or become infected again. This simulation gave us a valuable dataset to compare how effective quarantine measures are against limiting the spread of the virus.

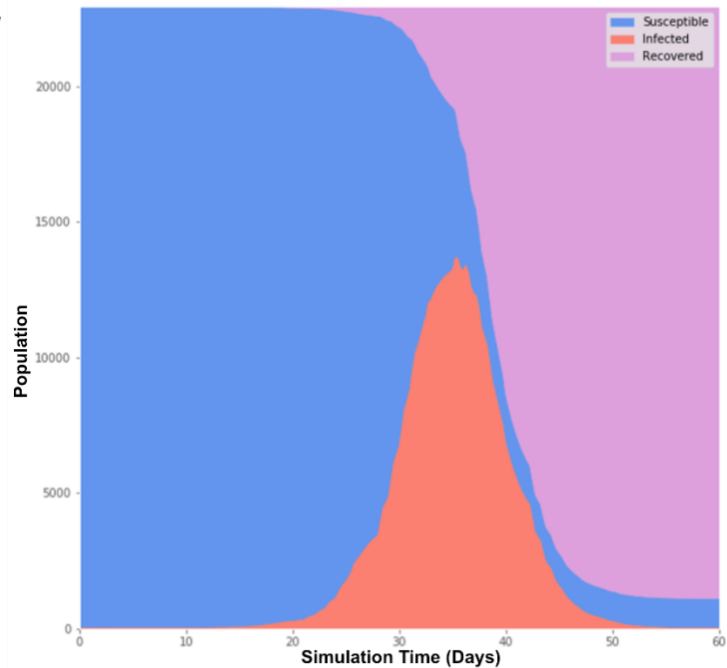

*Figure 7: SIR plot of generated by the model when no countermeasures are taken to prevent the spread of the virus.*

*B. Modeling quarantine measures and return to work*

After obtaining the baseline results, two sets of simulations were completed to measure the effectiveness of quarantining and returning to work at different levels of infection. In each case, when quarantine was initiated, all schools immediately transition to online classes. To simulate different levels of quarantine, one simulation for each set was run with 50% of the population working during quarantine and one simulation for each set has 10% of the population working during quarantine.

    *a. Quarantine at 5% infected, return to work at 1% infected*

The first set of quarantine simulations initiated a quarantine when 5% of the population became infected and had people return to their jobs when 1% of the population was infected. The black lines in the below Figures 8a and 8b illustrate the beginning and end of the quarantine.

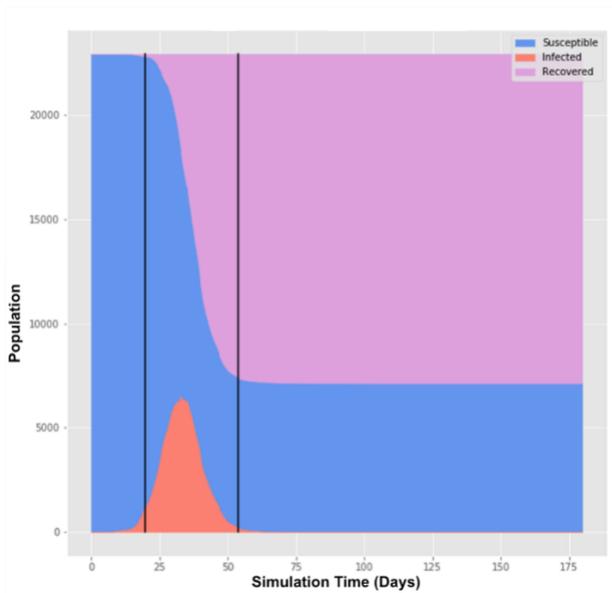 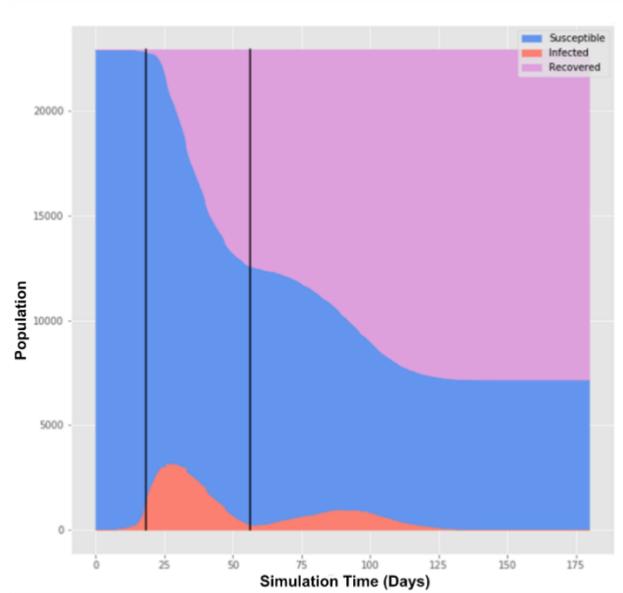

*Figure 8a: SIR plot generated by model when a quarantine is instituted when 5% of the population is infected. During quarantine 50% of the workforce still visits work in person each weekday. Quarantine is ended when the number of infected persons drops below 1% of the population.*

*Figure 8b: SIR plot generated by model when a quarantine is instituted when 5% of the population is infected. During quarantine 10% of the workforce still visits work in person each weekday. Quarantine is ended when the number of infected persons drops below 1% of the population.*

Figure 8a shows the results of the simulation when 50% of the population continues working during quarantine. At the peak of the wave 6700 people are infected. The next simulation, seen in Figure 8b, shows 10% of people still going to work during quarantine. The peak of the 1st wave occurs when 3600 people are infected. Approximately 1500 are infected at the peak of the 2nd wave.  As seen in Figure 2a, the peak number of infected is higher when 50% of people are working throughout quarantine.  When only 10% of people are working, the peak number of infected is lower; however, there is a second wave of increased infections after quarantine countermeasures are lifted.  Both methods ultimately result in nearly the same number of recovered/dead.

   b.  *Quarantine at 1% infected, return to work at 0.5% infected*

The following two simulations institute a quarantine when 1% of the population becomes infected and lifts the quarantine when 0.5% of the population is still infected.  However, two different levels of "essential workers" are used for these simulations to identify the effect limiting essential businesses has on the virus spread.  The first simulation was conducted with 50% of the population working during quarantine and the second simulation was conducted with 10% of the population working during quarantine.

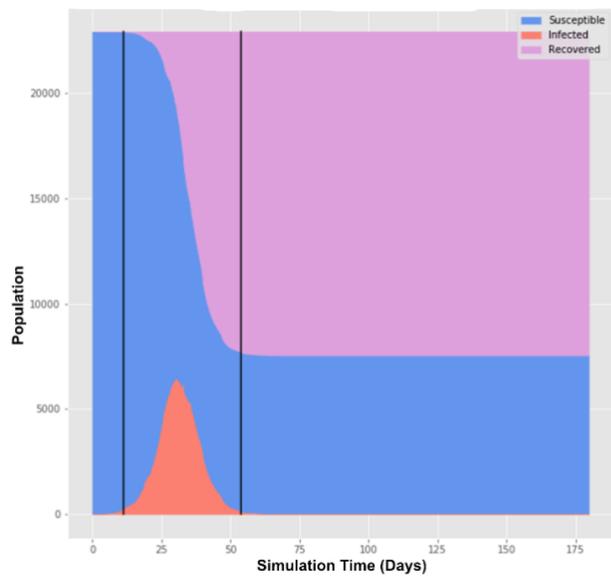 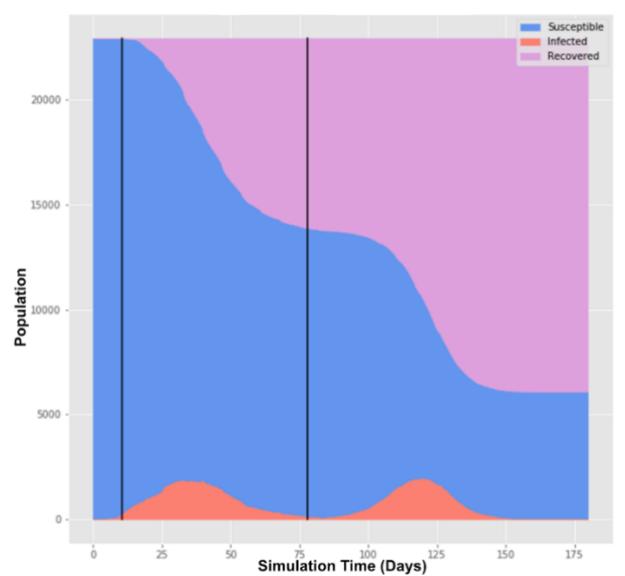

*Figure 9a: SIR plot generated by model when a quarantine is instituted when 1% of the population is infected. During quarantine 50% of the workforce still visits work in person each weekday. Quarantine is ended when the number of infected persons drops below 0.5% of the population.*

*Figure 9b: SIR plot generated by model when a quarantine is instituted when 1% of the population is infected. During quarantine 10% of the workforce still visits work in person each weekday. Quarantine is ended when the number of infected persons drops below 0.5% of the population.*

Figure 9a shows the simulation when 50% of the population continues to work. Approximately 6600 people become infected at the peak. Figure 9b shows that around 2000 people are infected at the peak of the first and second wave as 10% of people continue to work during quarantine. These peaks are both lower than the peak of the simulation where 50% of the population continues to work. The simulation with 10% people working has slightly more people within the "recovered" category.

*C. Modeling gradual and immediate return to work*

To measure the effectiveness of a gradual return to work over a sudden return to work, two simulations were conducted. In both simulations, quarantine was initiated at 5% infected and a work return was initiated at 1% infected. Additionally, 10% of the population worked during quarantine for both simulations. In the case of the immediate return to work, the population is sent back to work all at once. With a gradual return to work, workers resume in-person work gradually throughout a two-week period. The gradual return to work is designed to simulate different fields of work returning to work at different times.

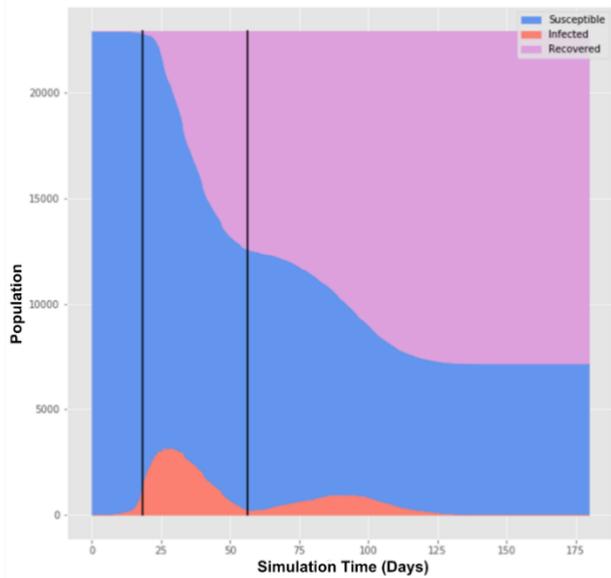 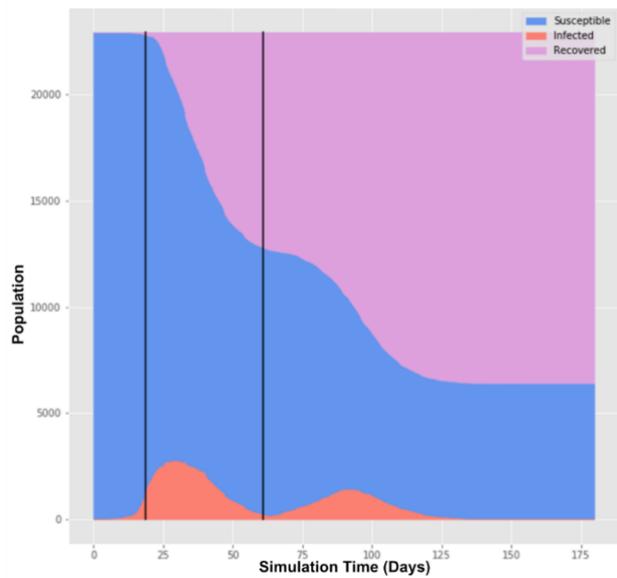

*Figure 10a: SIR plot generated by model when a quarantine is instituted when 5% of the population is infected. During quarantine 10% of the workforce still visits work in person each weekday. Quarantine is ended when the number of infected persons drops below 1% of the population. At the end quarantine the remainder of the workforce returns to in-person work in an immediate fashion*

*Figure 10b: SIR plot generated by model when a quarantine is instituted when 5% of the population is infected. During quarantine 10% of the workforce still visits work in person each weekday. Quarantine is ended when the number of infected persons drops below 1% of the population. At the end quarantine the remainder of the workforce returns to in-person work in a gradual manner over two weeks.*

Figure 10a shows the immediate return to work. The image on the right, Figure 10b, is for the gradual return to work. The initial peaks match each other in both cases, as is expected. In contrary, the second wave of infections differ due to the return to work process differences. The results of the model show that an immediate return to work resulted in less infections than a gradual return. This result initially surprised the modelers since it seems counter-intuitive. Our model calculates the probability of transmission as a function of the number of people at a given location. If there are fewer people at a location, they have a higher chance of interacting with each other and thus being in the same location as an infected person with a low number of total people results in a higher infection probability. More simulations still need to be run to fully explore this trend and determine its accuracy.

*D. Modeling the effect of panic buying due to quarantine announcement*

As can be readily seen at the onset of the Covid-19 pandemic in the United States, initial announcements of quarantines and restrictions often result in "panic" for a short period of time. During these panic periods, the population swarms grocery stores and supercenters to buy essential items for an extended stay in their home. A result of this panic is higher density and traffic in these supermarket stores. To determine the effects of these panic buying periods, two simulations were run. In both simulations, quarantine was initiated at 1% infected and a work return was initiated at 0.5% infected. One simulation implemented a three-day panic in which the odds an individual goes to a grocery store are increased drastically. The results of these simulations can be seen in Figure 11. While difficult to see in Figure 11, the panic buying results in almost 350 more cases of Covid-19 at the peak of infection. While 350 cases do not seem like many, this increase is for a scaled population of 23,000 people. For the entire Greater Cincinnati population of around 2 million people, this panic buying period could result in upwards of 30,000 additional Covid-19 positive cases.

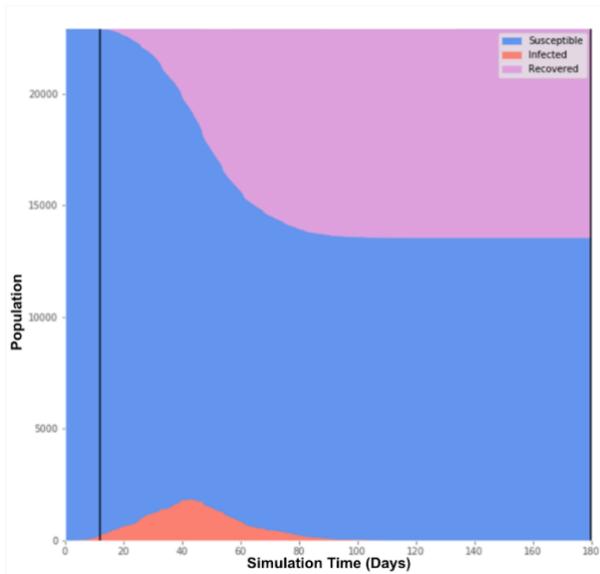

*Figure 11a: SIR plot generated by model when a quarantine is declared when 1% of the population is infected. Immediately after quarantine a 3 day "Panic" period is simulated. This period consists of highly rates of visiting grocery stores and supercenters. During quarantine 10% of the population continues to work in person. In this specific run quarantine is not ended because the first wave of infections was the focus.*

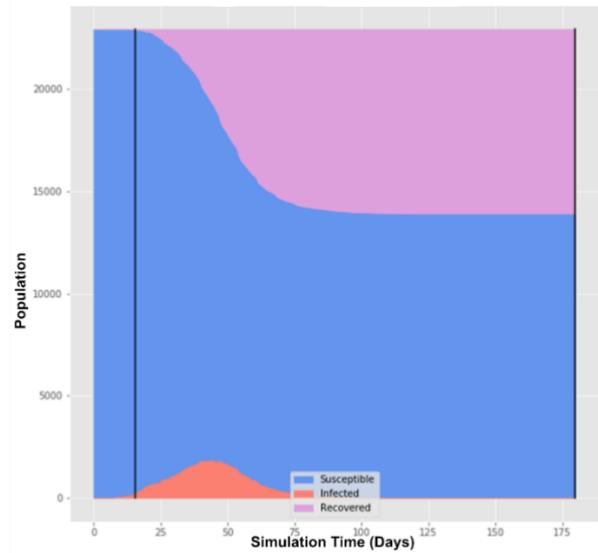

*Figure 11b: SIR plot generated by model when a quarantine is declared when 1% of the population is infected. In this run there is NO "Panic" period. During quarantine 10% of the population continues to work in person. In this specific run quarantine is not ended because the first wave of infections was the focus.*

| Figure Number | Description | Max Infected | Not Infected | Recovered/Removed |
|---|---|---|---|---|
| 7 | No countermeasures (baseline) | 14200 | 1100 | 21848 |
| 8a | Quarantine at 5% <br> Work return at 1% <br> 50% Work | 6766 | 7315 | 15633 |
| 8b | Quarantine at 5% <br> Work return at 1% <br> 10% Work | 3611 | 6983 | 15955 |
| 9a | Quarantine at 1% <br> Work return at 0.5% <br> 50% Work | 6621 | 7547 | 15401 |
| 9b | Quarantine at 1% <br> Work return at 0.5% <br> 10% Work | 2027 | 6081 | 16867 |
| 10a | Quarantine at 5% <br> Work return at 1% <br> 10% Work <br> Immediate return to work | 6766 | 7315 | 15633 |
| 10b | Quarantine at 5% <br> Work return at 1% <br> 10% Work <br> Gradual return to work | 2872 | 6542 | 16406 |
| 11a | Quarantine at 1% <br> Work return at 0.5% <br> 3 day panic | 1925 | 13571 | 9377 |
| 11b | Quarantine at 1% <br> Work return at 0.5% <br> No panic | 1961 | 13914 | 9034 |

*Table 1. Simulated final data for epidemic*

IV. DISCUSSION

The simulations conducted were aimed to measure three items:
- The effectiveness of quarantine
- The effect of a gradual return to work over an immediate return to work
- The effect of a sudden panic buying period when quarantine announcements are made

The model helped reinforce some of our predictions of the effectiveness of quarantine in the Tri-state area. First, instituting a quarantine lowered the total number of cases within the population. This is apparent by comparing the number of "Recovered" people in Figures 8 and 9 (with quarantine) with Figure 7 (no quarantine). Furthermore, our models show the amount of people working during quarantine plays a large role in the spread of the virus and the effects of a second wave. Comparing Figure 8a with 8b or Figure 9a with 9b, illustrates these trends. The runs where 50% of the population working during quarantine (Figure 8a and 9a), have a much higher number of people infected at the peak when compared to the runs when only 10% of the population worked (Figure 8b and 9b). It can be seen from the plots that when a large percentage of the population gets the virus during the first wave, there is little or no second wave. On the other hand, when quarantine is taken more seriously, with only 10% of the population continuing to work, the wave lasts much longer, and the population is more susceptible to a second wave. It is important to realize that the runs with 10% and 50% of the population had similar numbers of "removed" people at the end of the run. This shows the effectiveness of a quarantine at reducing the strain on the health system by lowering the maximum number of infections at the peak of the wave. Additionally, it shows that to lower the overall number of people infected, a strict quarantine would need to be instituted and held until there were no infectious cases remaining in the population. However, it can also be clearly seen by comparing our quarantine runs to our run without quarantine (Figure 7) that any amount of quarantine or social distancing reduces the number of infections.

Studying the gradual versus immediate return to work, turned out to be the most challenging aspect of our project. Due to the methodology used to calculate a person's individual likelihood of becoming infected (Equation 1), a given person has a higher probability of becoming infected when they are at the same location as an infected person and there are only a few people there vs when it is a crowded. As a result, when our model slowly starts re-introducing people to the workplace the number of employees at each location is rather small. This meant that our model predicted that a gradual return to work would yield a slightly higher second wave then an immediate return to work. This can be seen when comparing Figure 10a with Figure 10b. Furthermore, a gradual return to work resulted in more people having contracted the virus overall. This result seemed counter-intuitive so making changes to the model to account for workplaces implementing PPE and social distancing may be needed.

The model's analysis of a "panic" induced by the announcement of quarantine ended up being quite interesting. Figure 11 measures the effect of a sudden panic buying period compared with no panic. The three-day panic period results in more overall Covid-19 cases, due to the increased number of people being exposed to potentially infectious people while at grocery stores and supermarkets. When this increased number of infected was scaled to the full size of the Cincinnati Tri-State area, the model predicted approximately 30,000 extra cases due to the panic of the quarantine announcement.

V. ACKNOWLEDGEMENT

This work is supported by ONR STEM Grant # N000141912486. We thank the experts from Naval Surface Warfare Centers at Crane and Dahlgren for their feedback.